\documentclass[]{spie}  

\usepackage{amsmath,amsfonts,amssymb}
\usepackage{float}
\usepackage{graphicx}
\usepackage[colorlinks=true, allcolors=blue]{hyperref}
\usepackage{multirow}
\usepackage{caption}
\usepackage{siunitx}
\usepackage{array}
\newcommand{\PreserveBackslash}[1]{\let\temp=\\#1\let\\=\temp}
\newcolumntype{C}[1]{>{\PreserveBackslash\centering}p{#1}}
\newcolumntype{R}[1]{>{\PreserveBackslash\raggedleft}p{#1}}
\newcolumntype{L}[1]{>{\PreserveBackslash\raggedright}p{#1}}

\newcommand{\lamD}{$\lambda/D$ }

\title{Optical modeling for the evaluation of HOWFSC on embedded processors}

\author[a]{Kian Milani}
\author[b]{Ewan Douglas}
\author[c]{Leonid Pogorelyuk}
\author[d]{Christopher Mendillo}
\author[e]{Kerri Cahoy}
\author[e]{Nicholas Belsten}
\author[e]{Brandon Eickert}
\author{Shanti Rao}

\affil[a]{James C. Wyant College of Optical Sciences}
\affil[b]{Steward Observatory}
\affil[c]{Rensselaer Polytechnic Institute}
\affil[e]{Massachusetts Institute of Technology}
\affil[d]{University of Massachusetts Lowell}

\begin{document}
\maketitle


\begin{abstract}
The correction of quasi-static wavefront errors within a coronagraphic optical system will be a key challenge to overcome in order to directly image exoplanets in reflected light. These quasi-static errors are caused by mid to high-order surface errors on the optical elements as a result of manufacturing processes. Using high-order wavefront sensing and control (HOWFSC) techniques that do not introduce non-common path aberrations, the quasi-static errors can be corrected within the desired region of interest designated as the dark hole. For the future Habitable Worlds Observatory (HWO), HOWFSC algorithms will be key to attaining the desired contrasts. 

To simulate the performance of HOWFSC with space rated processors, optical models for a 6 m class space-borne observatory and a coronagraph have been developed. Phenomena such as the Talbot effect and beamwalk are included in the simulations using combinations of ray-based modeling and end-to-end propagation techniques. After integrating the optical models with the embedded processors, simulations with realistic computation times can be performed to understand the computational hardware performance that will be needed to maintain the desired contrasts. Here, the details of the optical models are presented along with the HOWFSC methods utilized. Initial results of the HOWFSC methods are also included as a demonstration of how system drifts will degrade the contrast and require dark hole maintenance. 
\end{abstract}

\section{Introduction}
At the recommendation of the Astro2020 decadal survey, NASA was advised to pursue a Habitable Worlds Observatory (HWO) equipped with a coronagraph capable of detecting exoplanets at 1E-10 contrast levels\cite{decadal_survey_on_astronomy_and_astrophysics_2020_astro2020_pathways_2021}. The Nancy Grace Roman Coronagraph Instrument will be a vital technology demonstration for the HWO coronagraph as it will utilize high-order wavefront sensing and control (HOWFSC) methods in order to reach contrasts on the order of 1E-8. To do so, the Roman Coronagraph will utilize a "set-and-forget" HOWFSC scheme involving ground-in-the-loop operations. However, this scheme will likely be unfeasible for the HWO coronagraph due to the higher sensitivity to drifts in the optical system at better contrasts. This means both reaching and maintaining the 1E-10 contrast will likely require HOWFSC iterations on time scales on the order of seconds to minutes\cite{pogorelyuk_computational_2022}. To accomplish this, a continuous HOWFSC scheme will have to be implemented to perform dark hole maintenance. This scheme will require embedded processors capable of performing the HOWFSC computations within the coherence time of the coronagraphic speckles. 

Because flight hardware can be "frozen" up to 10 years in advance of the mission launch, work has begun on implementing these HOWFSC algorithms on radiation hardened-processors to assess their performance, identify computational bottlenecks, and evaluate the impact on the performance of the instrument. To accomplish the third goal, we have implemented a framework for modeling a potential HWO coronagraph and simulating optical disturbances such as beamwalk. Here, the details of the optical models are presented while the methodology and implmentation of HOWFSC algorithms on space-rated processors is presented in Belsten et al\cite{belsten_evaluating_2023}. Currently, the only HOWFSC algorithms being investigated are pair-wise probing (PWP) and electric field conjugation (EFC)\cite{giveon_broadband_2007}. 

\section{Telescope Model}
To begin with, a nominal design for an off-axis three mirror anastigmat (TMA) telescope was created with Zemax to serve as a baseline for what a HWO could look like. This model is loosely based on the LUVOIR-B prescription in that it contains a similar net focal ratio of about F/36 and a similar primary mirror focal ratio of about F/3\cite{bolcar_large_2018}. However, the total pupil diameter has been shrunk from the 8.4m for LUVOIR-B to 6.5m. Additionally, an unobscured circular pupil is assumed unlike the hexagonally segmented pupil of LUVOIR-B. Note that this telescope is not completely optimized for diffraction limited imaging over a large FOV that will be desired for a HWO, but merely serves as a general architecture for the optical models required. As details about the HWO become available, this model will be updated to be more accurate. 

\begin{figure}[H]
    \centering
    \raisebox{-0.5\height}{\includegraphics[scale=0.25]{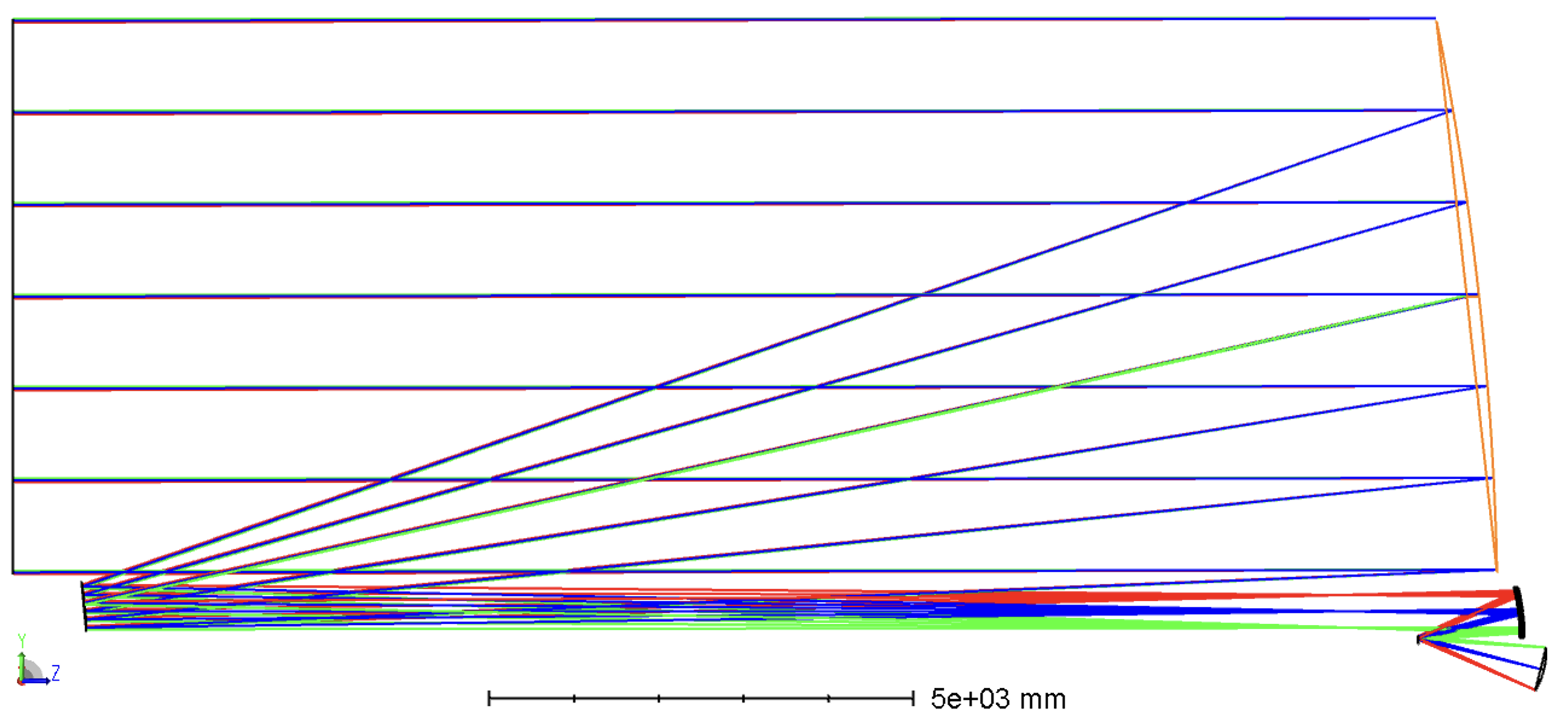}}
    \caption{This depicts the nominal design of the 6.5m class telescope currently being used for these optical models. The telescope is assumed to use a TMA architecture such that it can have a wider field of view to serve multiple instruments. Here, the field traced in green is designated as the coronagraph input as it has the smallest incident angles on the optics.}
    \label{fig:telescope}
\end{figure}

Using the raytrace model, a Fresnel model of the telescope is constructed using POPPY as the backend propagation software\cite{perrin-poppy-2016}. This model allows for the surface roughness of the telescope optics to be taken into account when performing the complete optical propagation including the coronagraph. But prior to including the WFE from each surface, the Fresnel model is validated by comparing the footprint diameter of each optic with the calculated footprint from the raytrace model. Additionally, the wavefront at M4 is analyzed to ensure it is a reimaged pupil as is expected from the Zemax design. Finally, the PSF of the Fresnel model is validated by computing the expected resolution of the telescope with the wavelength and F-number and comparing this resolution with the PSF result of the Fresnel model.  

\begin{figure}[H]
    \centering
    \raisebox{-0.5\height}{\includegraphics[scale=0.5]{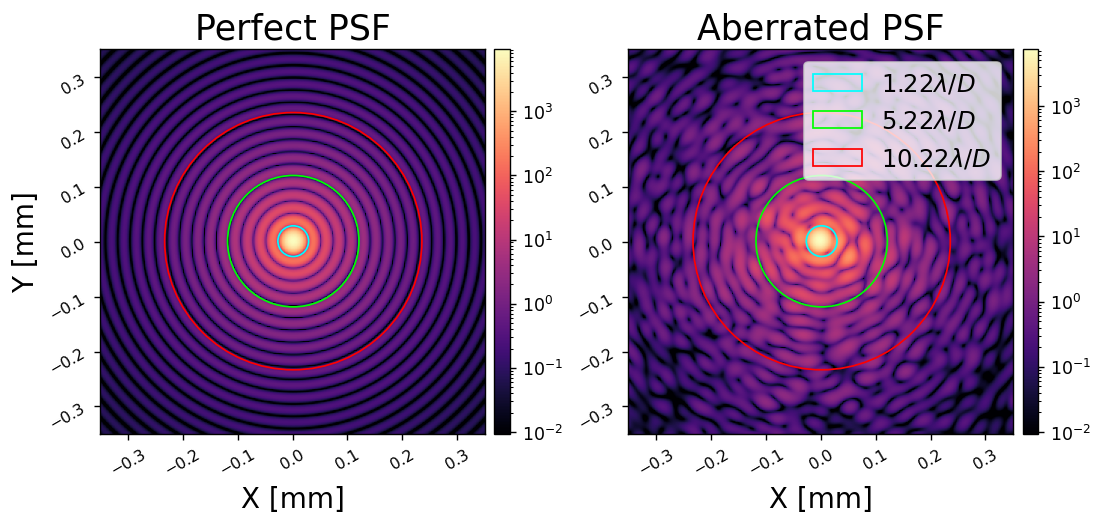}}
    \caption{Using POPPY, a Fresnel model of the telescope is constructed to propagate the effects from high-order surface roughness errors. Here, the image to the left is a PSF from the Fresnel model without the WFE contributions. The overlay of the circles at 1.22\lamD, 5.22\lamD, and 10.22\lamD demonstrate that the Fresnel model is in agreement with the expected resolution computed using the F-number from the raytrace model. The PSF to the right is the result after including each surface's WFE contribution.}
    \label{fig:tele-psfs}
\end{figure}

\begin{figure}[H]
    \centering
    \raisebox{-0.5\height}{\includegraphics[scale=0.5]{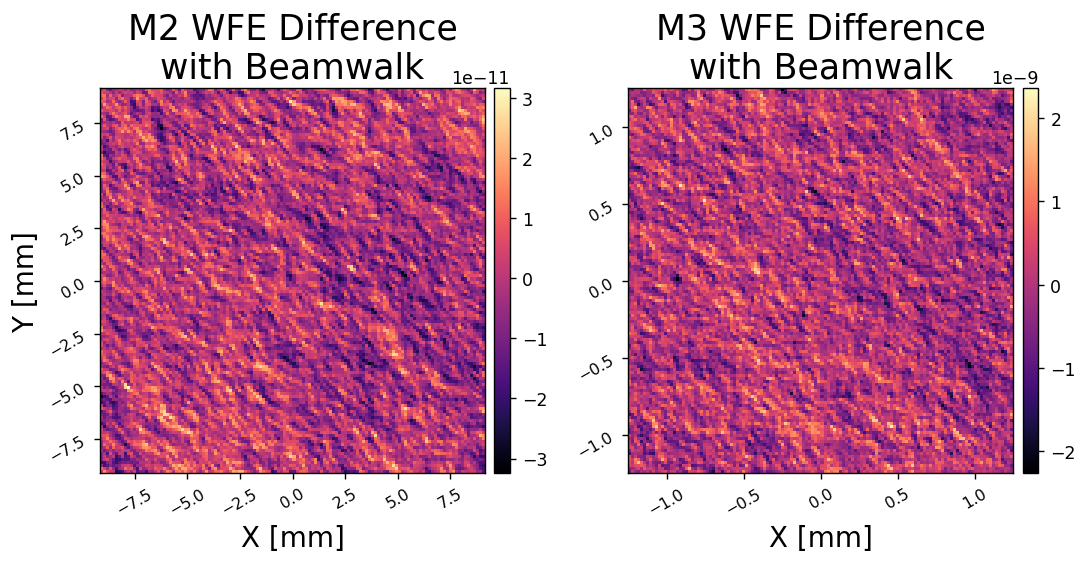}}
    \raisebox{-0.5\height}{\includegraphics[scale=0.5]{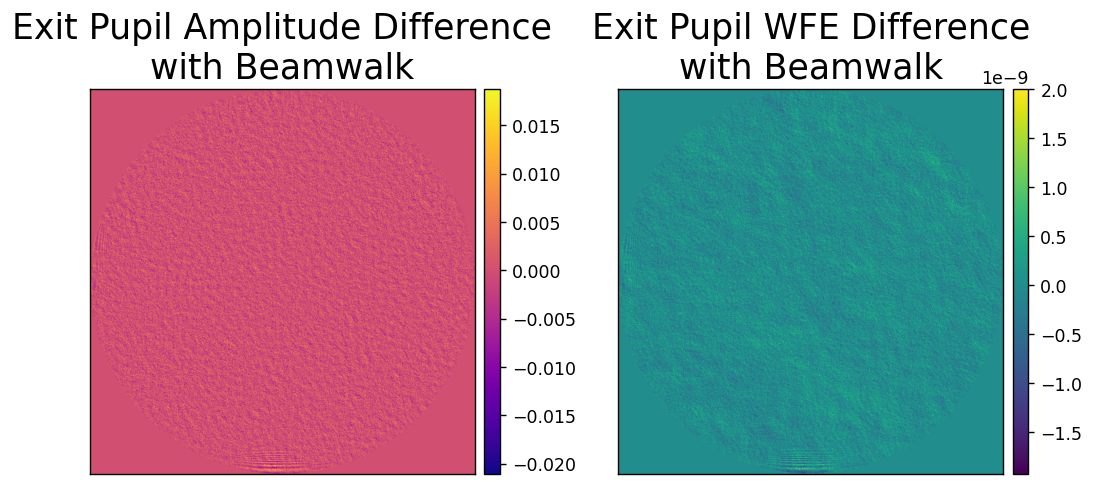}}
    \caption{Here, the difference in WFE on both M2 and M3 are are illustrated in the top row after computing the beamwalk induced by 15mas of pointing error. In the bottom row are the differences in both the exit pupil's amplitude and OPD after computing the effect of beamwalk and comparing with the nominal exit pupil wavefront (no pointing error).}
    \label{fig:bw-example}
\end{figure}

Using POPPY's StatisticalPSDWFE functionality, WFE maps are pre-computed for each optic in the telescope model. At the moment, the RMS WFE from the surface roughness of M1-M4 is 40nm, 20nm, 20nm and 15nm respectively. The current PSD of each surface is defined with a simple power law that has an index of -2.75, but this PSD will be updated in the future with more realistic parameters. These pre-computed WFEs are then used to simulate the effects of beamwalk on M2 and M3. Given M1 and M4 are pupils, beamwalk is not considered for these optics. Additionally, because M4 is a relayed pupil, we assume it can act as FSM such that beamwalk from pointing errors can be neglected on downstream optics. To implement the beamwalk, the footprint shift on M2 and M3 is assumed to be linear with respect to pointing error, so the Zemax model is used to compute the shift per milliarcsecond of pointing error. These values are found to be 0.084micron/mas and 0.908micron/mas for M2 and M3 respectively. Similar to Mendillo et al.\cite{mendillo_optical_2017}, the WFE maps for each surface are shifted by the appropriate values with subpixel precision. For this model, scipy.ndimage.shift (or the CuPy equivalent) are used. Figure \ref{fig:bw-example} illustrates the difference in WFE for each optic for 15mas of pointing error as well as the difference in the final pupil of the telescope computed by performing the propagation with and without the shifted WFEs. 

\section{Coronagraph Models}

To perform the HOWFSC experiments with a space-rated processor in the loop, additional Fresnel models have been created to simulate images from a "true" coronagraph. This Fresnel model also uses POPPY for the backend propagation and runs on a standard PC that will be used to compute the "true" images while the HOWFSC algorithm running on the space-rated processor will use the images to compute DM commands that are fed back to the coronagraph model. While computations are performed on the processor, system drifts can be simulated in both the telescope and coronagraph Fresnel models to evaluate the impact of compute times. At the moment, there are no relay optics included in the design because beamwalk or other drifts from these optics are assumed to be negligible given they are after the FSM/M4. This allows the coroangraph model to be slightly simplified such that the wavefront of the telescope exit pupil is computed and directly injected into the coronagraph model rather than being propagated through relay optics. 

\begin{figure}[H]
    \centering
    \raisebox{-0.5\height}{\includegraphics[scale=0.35]{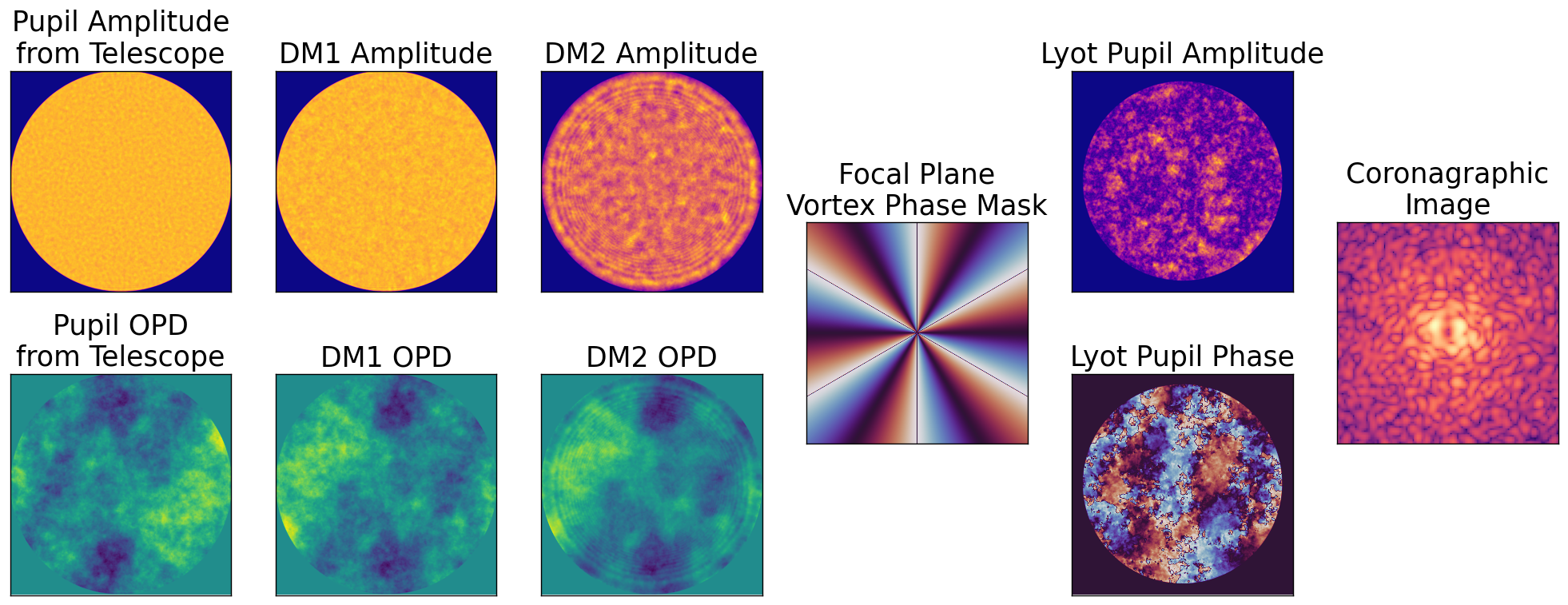}}
    \caption{This figure depicts the main optical components being used for the vortex coronagraph. The input WFE at the entrance pupil of the coronagraph is computed using the Fresnel model of the telescope and injected into the coronagraph model.}
    \label{fig:coro}
\end{figure}

Here, only a simple vortex coronagraph is considered as VVCs have previously been considered for a HabEx mission\cite{krist_numerical_2019}. Because our telescope model does not include any segmentation, the coronagraph model does not include any apodization or DM assistance for the vortex, although, an additional pupil plane where an apodizer may be placed is included so the model may be updated in the future. Figure \ref{fig:coro} illustrates the fundamental optical train of the coronagraph model. Here, the deformable mirrors (DMs) are being modeled using the fast convolution method described in Will et al.\cite{will_wavefront_2021}. To numerically model a vortex phase mask, the same method described in Krist et al\cite{krist_numerical_2019} is being used. When implementing this in a Fresnel model using POPPY, the model is separated into two segments. The first propagates from the entrance pupil of the coronagraph to the focal plane where the vortex will be located. At this point, the propagation through POPPY is ended and an FFT is used to compute the pupil plane wavefront from the focal plane data. Now, the vortex is numerically applied with additional FFTs and MFTs which output the wavefront at a pupil plane after the vortex. Angular spectrum propagation is then used to back propagate the pupil wavefront to an OAP that would collimate the beam coming from the vortex mask. A surface roughness map is applied to this wavefront and angular spectrum is used to propagate back to the pupil. This data now acts as the wavefront at the Lyot stop, so POPPY is again used to propagate from the Lyot stop through the rest of the optical train and to the image plane. 

\begin{figure}[H]
    \centering
    \raisebox{-0.5\height}{\includegraphics[scale=0.45]{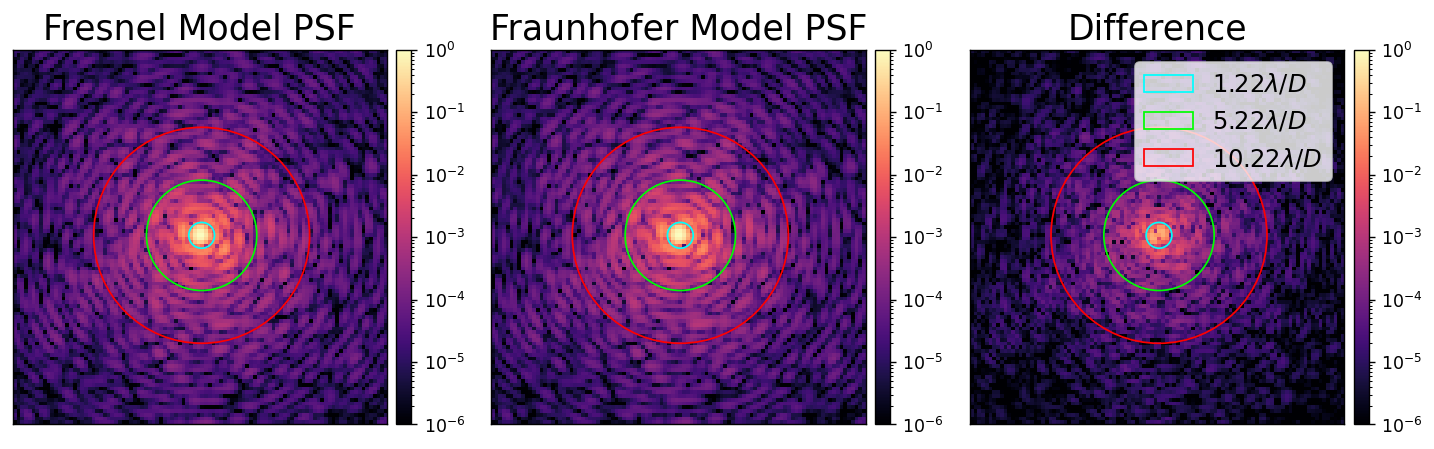}}
    \raisebox{-0.5\height}{\includegraphics[scale=0.45]{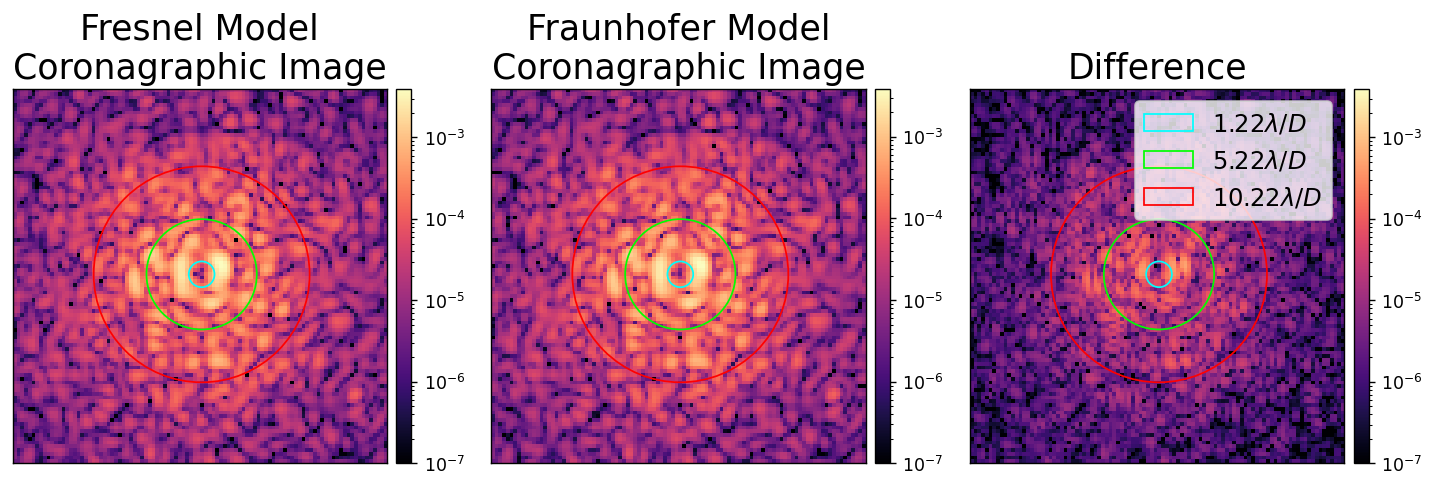}}
    \caption{Here, the images in the top demonstrate a comparison of the PSF from the Fresnel model of the coronagraph and the PSF from the Fraunhofer model. In the bottom row is a similar comparison with coronagraphic images of each model. The morphology of the PSFs and coronagraphic images demonstrate the agreement between the two models. }
    \label{fig:fvf}
\end{figure}

With the Fresnel model acting as the true coronagraph, a compact/Fraunhofer model is also created using FFTs and MFTs for the implementation of HOWFSC algorithms. The pre-vortex WFE is then computed with the Fresnel model and injected into the compact model. In reality, phase retrieval techniques would be used to measure the WFE of an instrument and inject the measurement into the model, but no phase retrieval method has currently been implemented. Figure \ref{fig:fvf} presents a comparison of the Fresnel model PSFs and coronagraphic images with the injected WFE. Because the pre-FPM errors are more significant to the coronagraph, the residual surface errors after the FPM are ignored in this compact model. Nonetheless, the morphology of the PSF and speckles in the compact model demonstrates agreement with the Fresnel model, so it acts as a well calibrated model for HOWFSC. 

\begin{figure}[H]
    \centering
    \raisebox{-0.5\height}{\includegraphics[scale=0.5]{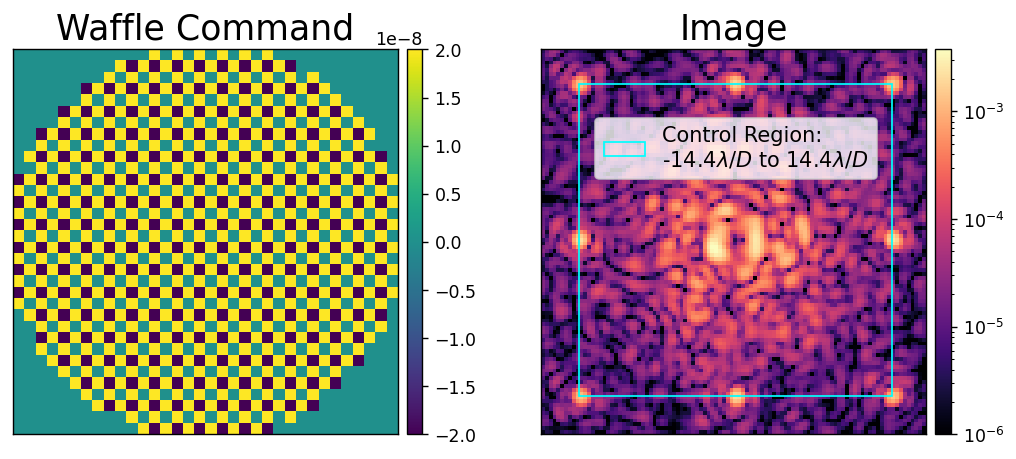}}
    \raisebox{-0.5\height}{\includegraphics[scale=0.5]{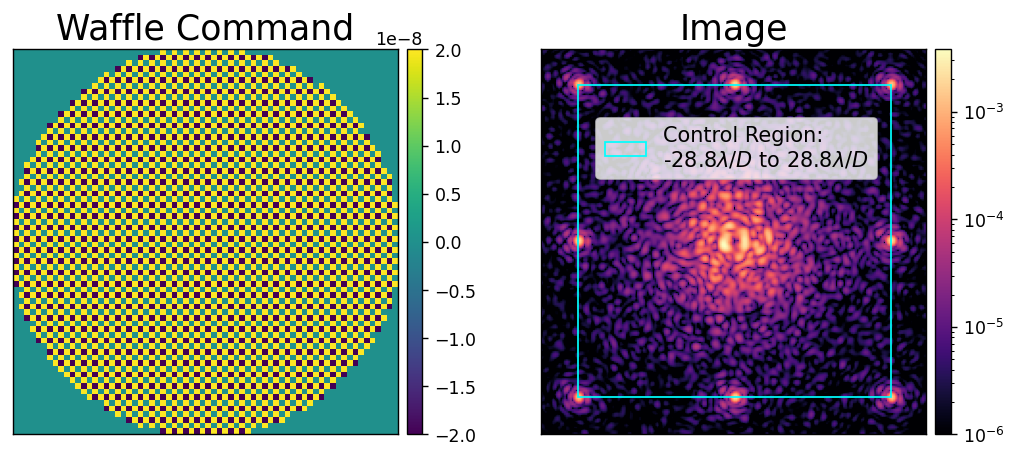}}
    \caption{The images here demonstrate the control regions for each DM configuration being considered using a waffle command to produce speckles at the edges of each potential control region. Note that here, $D$ is the diameter of Lyot stop, which is undersized from the geometric pupil diameter.}
    \label{fig:waffles}
\end{figure}

For most HOWFSC algorithms, the computational complexity will be dependent on the number of actuators (or DM modes) being utilized and the number of pixels in the focal plane within the desired control region. To evaluate the performance of processors for varying actuator counts, two configurations of this coronagraph model are created. The first assumes smaller 34x34 actuator DMs while the second uses 68x68 actuator DMs. The larger DM model assumes the same actuator spacing, but the pupil diameter is doubled to account for the higher actuator count. The detector sampling in each model is assumed to be 5 microns, but the final imaging focal length is also doubled for the model with larger pupils such that the pixelscale is 0.354\lamD in each. Table \ref{tab:dm-params} contains the details of each models pupil sizes and actuator counts while Figure \ref{fig:waffles} illustrates the difference in potential control regions with the higher actuator count. Both models will be used when evaluating the processors to understand how the performance will scale with more actuators and larger control regions. 

\begin{table}[H]
    \centering
    \caption{The assumptions for the two DM configurations being considered for the HOWFSC implementation are presented here.}
    \begin{tabular}{|L{4cm}|L{3cm}|L{3cm}|} \hline
          & Small DM model & Large DM model \\ \hline
        Actuators across DM  & 34 & 68 \\ \hline
        Total Actuator Count  & $2\times952$ & $2\times3720$ \\ \hline
        Actuator Spacing  & 0.3mm & 0.3mm \\ \hline
        Pupil Size at DM  & 9.6mm & 19.2mm \\ \hline
        Actuators Across Pupil  & 32.0 & 64.0 \\ \hline
    \end{tabular}
    \label{tab:dm-params}
\end{table}

\section{HOWFSC Simulations}
Currently, the primary HOWFSC methods being considered for dark hole creation and maintenance are standard EFC and PWP. Two additional HOWFSC methods will be implemented in the future including modal PWP demonstrated by Pogorelyuk et al.\cite{pogorelyuk_dark_2022} for more efficient dark hole maintenance along with the Jacobian-free algorithmic differentiation EFC introduced by Will et al\cite{will_jacobian-free_2021}. For now, the models discussed above are used to perform EFC and illustrate why dark hole maintenance will be necessary, particularly for the 1E-10 contrasts. 



\begin{figure}[H]
    \centering
    \raisebox{-0.5\height}{\includegraphics[scale=0.4]{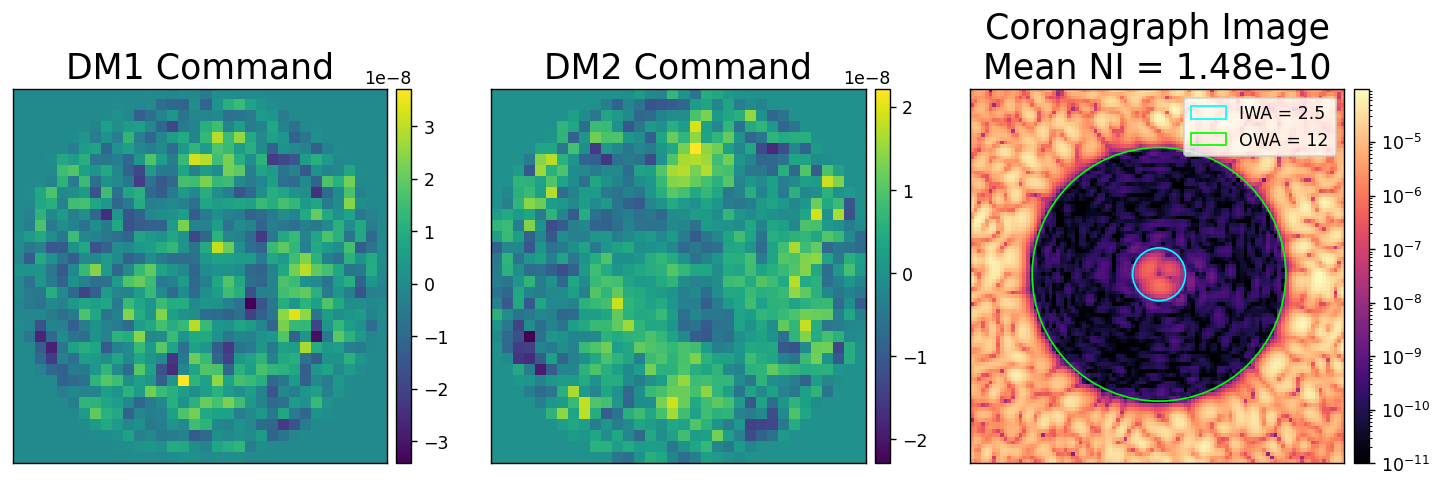}}
    \raisebox{-0.5\height}{\includegraphics[scale=0.4]{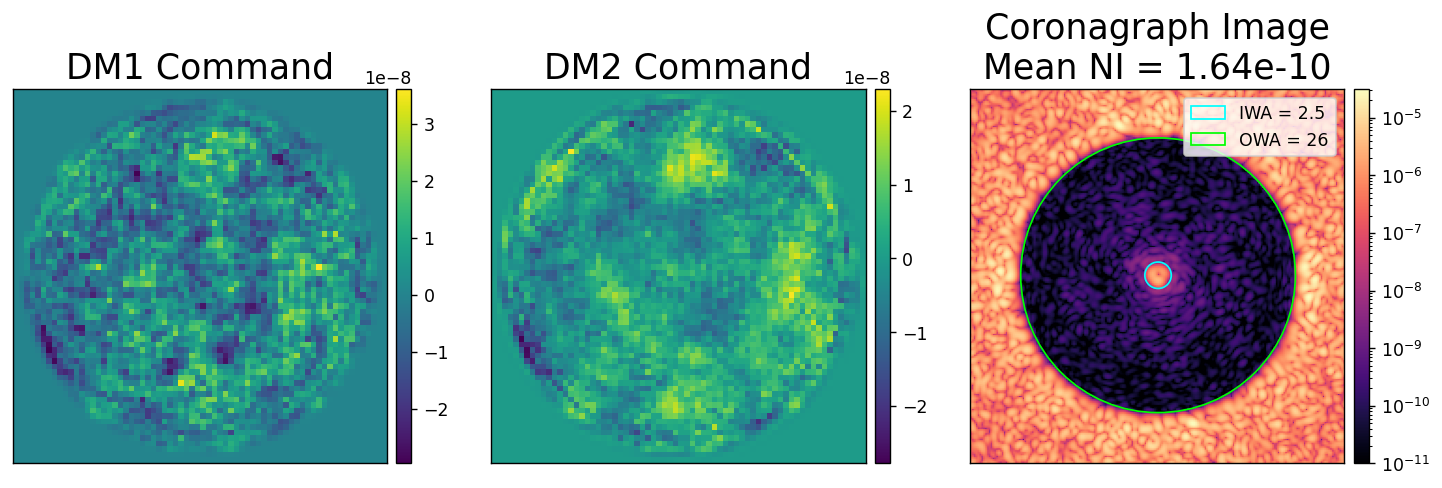}}
    \caption{Using the Fraunhofer model of for each DM configuration of the coronagraph, EFC is used to produce a dark hole. In the top row are the solutions for the 34x34 DMs and the bottom row are the solutions with the 68x68 DMs. Each of the solutions are generated after 18 iterations of EFC using a single relinearization after 9 iterations.}
    \label{fig:efc}
\end{figure}

\begin{figure}[H]
    \centering
    \raisebox{-0.5\height}{\includegraphics[scale=0.45]{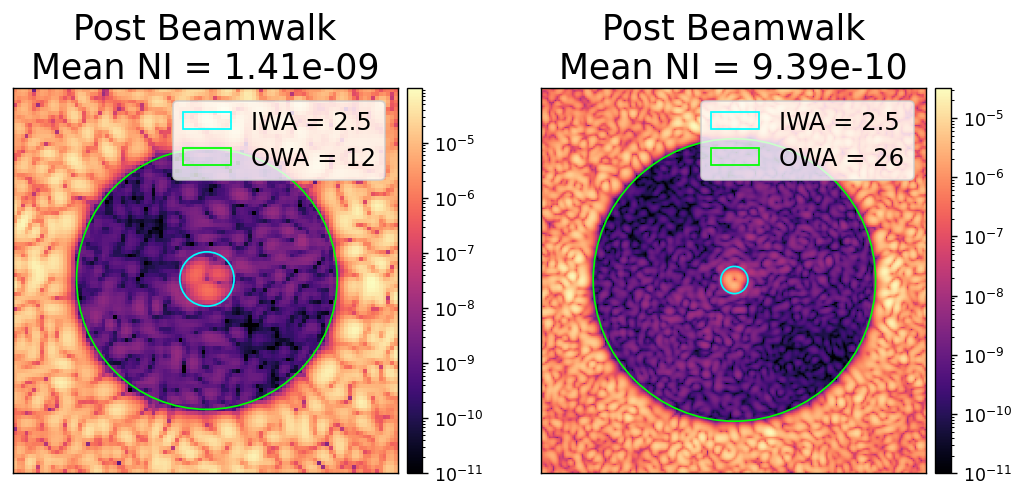}}
    \raisebox{-0.5\height}{\includegraphics[scale=0.45]{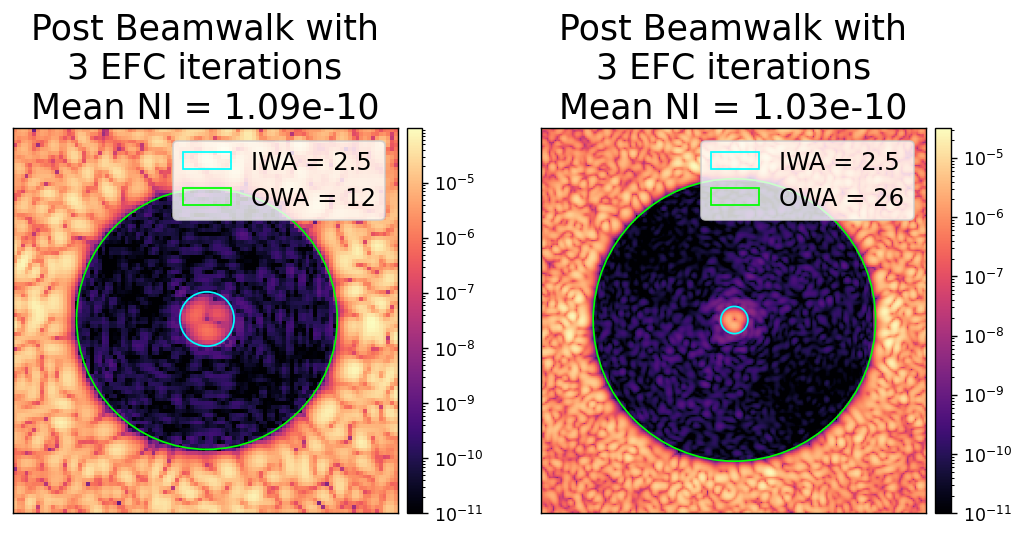}}
    \caption{In the top row are the images from each coronagraph model if we assume the telescope pointing drifts by 15mas (x + y). Here, the contrast in each dark hole degrades by about 1E-9 solely from the beamwalk on M2 and M3. In the bottom row, 3 additional iterations of EFC are used to correct the drift in the speckles and re-obtain the 1E-10 contrast with each DM configuration.}
    \label{fig:efc-beamwalk}
\end{figure}

For simplicity, only monochromatic EFC is implemented, but this will be expanded to broadband EFC in the future. Figures \ref{fig:efc} and \ref{fig:efc-beamwalk} illustrate an example of why using a HOWFSC method for dark hole maintenance will likely be necessary. Here, EFC is originally used to create a dark hole assuming a perfectly stable telescope. This allows us to generate a dark hole with 1E-10 contrast in just 18 iterations. One noteworthy point is that due to the large WFEs assumed in the telescope optics, the Jacobian was relinearized after 9 iterations in order to reach the final contrasts indicated in the figures. After the initial EFC loop is completed, a pointing error of 15mas was injected into the telescope model. The effect of beamwalk on M2 and M3 was then computed and injected into the coronagraph model to assess the contrast degradation this level of pointing error would induce assuming no other dynamics in the coronagraph system. As a result of this beamwalk, the contrast degrades by about an order magnitude with each DM configuration. However, the new speckles can be corrected with additional iterations of EFC. For the models being used here, each DM configuration utilized 3 more EFC iteration to re-converge to 1E-10 contrast. It should be noted that contrast degradation from beamwalk and other drifts in the coronagraph system will also depend on the quality of each optic's surface roughness as the better each optic can be polished, then larger drifts can be tolerated. 

\section{Conclusions and Future Work}

While a set-and-forget HOWFSC scheme is possible for more moderate contrast goals, the extreme contrast requirement of the HWO will require much more frequent dark hole maintenance with HOWFSC techniques due to drifts within the optical system. These drifts will have to be corrected on time scales equivalent to or smaller than the lifetime of the speckles in order for the HOWFSC maintenance to operate as intended. Here a framework has been developed to model various configurations and parameters of a 6.5m class telescope and coronagraph to evaluate the impact of HOWFSC computation times. As more details become available about potential HWO concepts including telescope design, coronagraph modes, and quality of optical surfaces, the models here will be updated to include the more accurate parameters. Future experiments with space-rated processors in-the-loop will also implement various other dynamics such as DM creep and slow shifts of coronagraph optics. These experiments will yield results informing the mission about how fast HOWFSC will need to run for various conditions and what the computational bottlenecks for HOWFSC will be. 

\section{Acknowledgements}
This work is supported by the NASA Astrophysics Technology Division under APRA grant \#80NSSC22K1412. This research made use of community-developed core Python packages, including: POPPY\cite{perrin-poppy-2016}, Astropy \cite{the_astropy_collaboration_astropy_2013}, Matplotlib \cite{hunter_matplotlib_2007}, SciPy \cite{jones_scipy_2001}, CuPy\cite{Okuta2017CuPyA}, Ray\cite{moritz_ray_2018}, and the IPython Interactive Computing architecture \cite{perez_ipython_2007}.

\bibliographystyle{spiebib} 
\bibliography{spie2024, citations}

\begin{thebibliography}{10}

\bibitem{decadal_survey_on_astronomy_and_astrophysics_2020_astro2020_pathways_2021}
{Decadal Survey on Astronomy and Astrophysics 2020 (Astro2020)}, {Space Studies Board}, {Board on Physics and Astronomy}, {Division on Engineering and Physical Sciences}, and {National Academies of Sciences, Engineering, and Medicine},  [{\em Pathways to Discovery in Astronomy and Astrophysics for the 2020s}{\nolinebreak\hspace{0.1em}]}, National Academies Press.
\newblock Pages: 26141.

\bibitem{pogorelyuk_computational_2022}
Pogorelyuk, L., Haughwout, C., Belsten, N., Cady, E., and Cahoy, K., ``Computational complexities of image plane algorithms for high contrast imaging in space telescopes,'' ~{\bf 8}(4).

\bibitem{belsten_evaluating_2023}
Belsten, N., Milani, K., Pogorelyuk, L., Eickert, B., Rao, S., Douglas, E.~S., and Cahoy, K., ``Evaluating embedded hardware for high-order wavefront sensing and control,'' in [{\em Techniques and Instrumentation for Detection of Exoplanets {XI}}{\nolinebreak\hspace{0.1em}]},  Ruane, G.~J., ed.,  60, {SPIE}.

\bibitem{giveon_broadband_2007}
Give'on, A., Kern, B., Shaklan, S., Moody, D.~C., and Pueyo, L., ``Broadband wavefront correction algorithm for high-contrast imaging systems,''  66910A.

\bibitem{bolcar_large_2018}
Bolcar, M.~R., Hylan, J.~E., Crooke, J.~A., Bronke, G., Collins, C., Corsetti, J.~A., Generie, J., Gong, Q., Groff, T., Hayden, W., Jones, A., Matonak, B., Park, S., Sacks, L., West, G., Yang, K., and Zimmerman, N.~T., ``The large {UV}/optical/infrared surveyor ({LUVOIR}): decadal mission study update,'' in [{\em Space Telescopes and Instrumentation 2018: Optical, Infrared, and Millimeter Wave}{\nolinebreak\hspace{0.1em}]},  {MacEwen}, H.~A., Lystrup, M., Fazio, G.~G., Batalha, N., Tong, E.~C., and Siegler, N., eds.,  23, {SPIE}.

\bibitem{perrin-poppy-2016}
Perrin, M., Long, J., Douglas, E., Zimmerman, N., Sivaramakrishnan, A., Douglass, K., and Grochowicz, M., ``Physical optics propagation in python,'' (2016).

\bibitem{mendillo_optical_2017}
Mendillo, C.~B., Howe, G.~A., Hewawasam, K., Martel, J., Finn, S.~C., Cook, T.~A., and Chakrabarti, S., ``Optical tolerances for the {PICTURE}-c mission: error budget for electric field conjugation, beam walk, surface scatter, and polarization aberration,'' in [{\em Techniques and Instrumentation for Detection of Exoplanets {VIII}}{\nolinebreak\hspace{0.1em}]},  Shaklan, S., ed.,  34, {SPIE}.

\bibitem{krist_numerical_2019}
Krist, J.~E., Martin, S.~R., Kuan, G., Mennesson, B., Ruane, G., Saini, N., Trauger, J., Breckinridge, J., Mawet, D., Stahl, P., and Davis, J., ``Numerical modeling of the habex coronagraph,'' in [{\em Techniques and Instrumentation for Detection of Exoplanets {IX}}{\nolinebreak\hspace{0.1em}]},  Shaklan, S.~B., ed.,  5, {SPIE}.

\bibitem{will_wavefront_2021}
Will, S.~D., Perrin, M.~D., Por, E.~H., Noss, J., Sahoo, A., Petrone, P., Laginja, I., Pourcelot, R., Redmond, S.~M., Pueyo, L., Groff, T.~D., Fienup, J.~R., and Soummer, R., ``Wavefront control with algorithmic differentiation on the {HiCAT} testbed,'' in [{\em Techniques and Instrumentation for Detection of Exoplanets X}{\nolinebreak\hspace{0.1em}]},  Shaklan, S.~B. and Ruane, G.~J., eds.,  28, {SPIE}.

\bibitem{pogorelyuk_dark_2022}
Pogorelyuk, L., Krist, J., Nemati, B., Riggs, A. J.~E., Miller, S., Pueyo, L., Kasdin, N.~J., and Cahoy, K., ``Dark hole maintenance with modal pairwise probing in numerical simulations of roman coronagraph instrument,'' ~{\bf 8}(1).

\bibitem{will_jacobian-free_2021}
Will, S.~D., Groff, T.~D., and Fienup, J.~R., ``Jacobian-free coronagraphic wavefront control using nonlinear optimization,'' ~{\bf 7}(1).

\bibitem{the_astropy_collaboration_astropy_2013}
{The Astropy Collaboration}, e.~a., ``Astropy: {A} community {Python} package for astronomy,'' {\em Astronomy \& Astrophysics}~{\bf 558},  A33 (Oct. 2013).

\bibitem{hunter_matplotlib_2007}
Hunter, J.~D., ``Matplotlib: {A} {2D} graphics environment,'' {\em Computing In Science \& Engineering}~{\bf 9}(3),  90--95 (2007).

\bibitem{jones_scipy_2001}
Jones, E., Oliphant, T., and Peterson, P., ``{SciPy}: {Open} source scientific tools for {Python},'' {\em http://www. scipy. org/}  (2001).

\bibitem{Okuta2017CuPyA}
Okuta, R., Unno, Y., Nishino, D., Hido, S., and Crissman, ``Cupy : A numpy-compatible library for nvidia gpu calculations,'' (2017).

\bibitem{moritz_ray_2018}
Moritz, P., Nishihara, R., Wang, S., Tumanov, A., Liaw, R., Liang, E., Elibol, M., Yang, Z., Paul, W., Jordan, M.~I., and Stoica, I., ``Ray: {A} {Distributed} {Framework} for {Emerging} {AI} {Applications},'' (Sept. 2018).
\newblock arXiv:1712.05889 [cs, stat].

\bibitem{perez_ipython_2007}
P{\'e}rez, F. and Granger, B., ``{IPython}: {A} {System} for {Interactive} {Scientific} {Computing},'' {\em Computing in Science Engineering}~{\bf 9},  21--29 (May 2007).

\end{thebibliography}

\end{document}